\documentclass{ws-ijmpd}

\begin{document}

\markboth{Partha Sarathi Pal, Sandip K. Chakrabarti, Anuj Nandi}
{Accretion flow geometry in GRS 1915+105}

\catchline{}{}{}{}{}
%
%%%%%%%%%%%%%%%%%%%%%%%%%%%%%%%%%%%%%%%%%%%%%%%%%%%%%%%%%%%%%%%%%%%%

\title{Evidence of variation of the accretion flow geometry
in GRS 1915+105 from IXAE and RXTE data}

\author{Partha Sarathi Pal$^1$, Sandip K. Chakrabarti$^{2,1}$ and Anuj Nandi$^{1,3}$}

\address{1. Indian Centre for Space Physics, Chalantika 43, Garia Station Rd. Kolkata 700084\\ 
2. S.N. Bose National Centre for Basic Sciences, JD Block, Kolkata 700098; chakraba@bose.res.in\\
3. Space Science Division, ISRO-HQ, New BEL Rd, Bangalore, 560231\\ }

\maketitle

%\begin{history}
%\received{Day Month Year}
%\revised{Day Month Year}
%\accepted{Day Month Year}
%\comby{(xxxxxxxxxx)}
%\end{history}

\begin{abstract}
The Galactic microquasar GRS 1915+105 exhibits various types of light curves.
There is, however, no understanding of when a certain type of light curve will
be exhibited and only in a handful of cases, the transitions from one
type to another have actually been observed.
We study the detailed spectral properties in these cases to show that
that different classes have different ratio
of the power-law photon and the blackbody photon. Since the power-law photons
are from the Compton cloud, and the intensity of the power-law photon component
depends on the degree of interception of the soft photons by the Compton cloud,
we conclude that not only the accretion rate, but the accretion flow
geometry must also change during a class transition.
\end{abstract}


\section{Introduction}	
Indian X-ray Astronomy Experiments (IXAE) payload on board IRS-P3 satellite data of the
enigmatic stellar mass black hole binary GRS 1915+105 (Harlaftis \& Greiner, 2004)
clearly indicated that GRS 1915+105 can have various types of light curves
(Yadav et al., 1999; Rao, Yadav \& Paul, 2000; Chakrabarti \& Nandi, 2000 and references therein;
Naik et al. 2002a). So far, a total of thirteen types of light 
curves have been detected. However, there are only a handful
of cases when an actual transition from one type to another has been
reported (Chakrabarti et al., 2004; Chakrabarti et al. 2005, hereafter Paper I and Paper II respectively),
and those too using the IXAE data which had only two energy channels.
It was possible to see the transitions because
the IXAE payload observed the same object, namely, GRS 1915+105, for several orbits.
Specifically, Papers I and II showed that in a matter of two to three
hours indicating that it is not the viscous time scale of the Keplerian disk, but the free-fall
time scale of the low-angular momentum halo which decides the variability class transition.

In the present paper, we study the spectra of the classes which showed the transitions and
found that one of the parameters which could be responsible for the class transition is the
geometry of the Compton cloud. There are various models of this cloud, including
a Keplerian  disk with corona (Haardt \& Maraschi, 1993) and the
unstable inner edge of the standard disk (Kobayashi et al. 2003). However, we shall
use a physically reasonable solution of the two component advective flow (TCAF) model
(Chakrabarti \& Titarchuk, 1995, hereafter CT95), where, the Compton
cloud is made up of the post-shock region of the sub-Keplerian component
which includes the jet and the outflows as well. In the literature many works are present 
which discusses the quasi-periodic oscillations of black hole candidates
(e.g., Strohmayer, 2001; Wagoner, Silbergleit,  Ortega-Rodríguez, 2001; Stella \& Vietri, 1999;
Abramowicz, Kluzniak, 2001; Stuchlík, Slaný, Török, 2007; Kato, 2008; Blaes, Sramkova, Abramowicz, 
Kluzniak \& Torkelson, 2007). However, we shall concentrate only on the spectral properties
of GRS 1915+105 in the current paper.

The plan of this {\it letter} is the following: we briefly present the data analysis procedure
in our next section. In \S 3, we study the nature of the components of
the spectra, especially the black body and the power-law components. We compute the
efficiency of Comptonization by taking the ratio of the photon numbers in those
components. We show that indeed, the average ratio varies from class to class.
Finally, in \S 4, we discuss the physics behind such changes
in relation to the accretion flow dynamics.

\section{Observation \& Spectral Analysis}
In Papers I and II we have already presented the IXAE data analysis procedure. Due to poor spectral
and temporal resolutions of IXAE, we take resort to the RXTE science data, which are taken from the NASA
HEASARC data archive for analysis. We chose the data procured in 1996-97 by RXTE as in this
period GRS 1915+105 has shown almost all types of variabilities in X-rays.
We however exclude the data collected for elevation angles less than $10^{\circ}$,
offset greater than $0.02^{\circ}$ and during the South Atlantic Anomaly (SAA) passage. Since
among the four PCUs only PCU1 and PCU2 were working most efficiently and are continuously ON through
the observations, we took the data from PCU1 and PCU2 with all the layers in our analysis. In Table. 1, the ObsIDs are given.

\begin{table*}
\begin{center}
%\begin{tabular}{crr}
\begin{tabular}{ccc}
Obs-Id & Class & Date \\
\hline
20402-01-35-00$^*$ &$\kappa$ &07-07-1997\\
20402-01-33-00 &$\kappa$ &18-06-1997\\
20402-01-31-00$^*$ &$\rho$ &03-06-1997\\
20402-01-03-00 &$\rho$ &19-11-1996\\
20187-02-01-00$^*$ &$\alpha$ &07-05-1997\\
20402-01-30-01 &$\alpha$ &28-05-1997\\
10408-01-15-00$^*$ &$\theta$ &16-06-1996\\
20402-01-45-02 &$\theta$ &05-09-1997\\
20402-01-16-00$^*$ &$\chi$ &22-02-1997\\
20402-01-05-00 &$\chi$ &04-12-1996\\
\end{tabular}
\end{center}
\caption{The ObsIDs and the dates of RXTE data which were analyzed in this paper.
$^*$ represents the result shown in Fig. 2 of this paper.}
\end{table*}

\begin{figure}[h]
\centerline{
\vbox{
\psfig{file=fig1.ps,angle=270.0,width=10truecm}}}
\noindent{\small{\bf Fig 1:} 
A sample spectrum with the fitted {\it diskbb} and power law components.
The HEXTE data is also added.}
\end{figure}

\begin{figure}[h]
\centerline{
\vbox{
\psfig{file=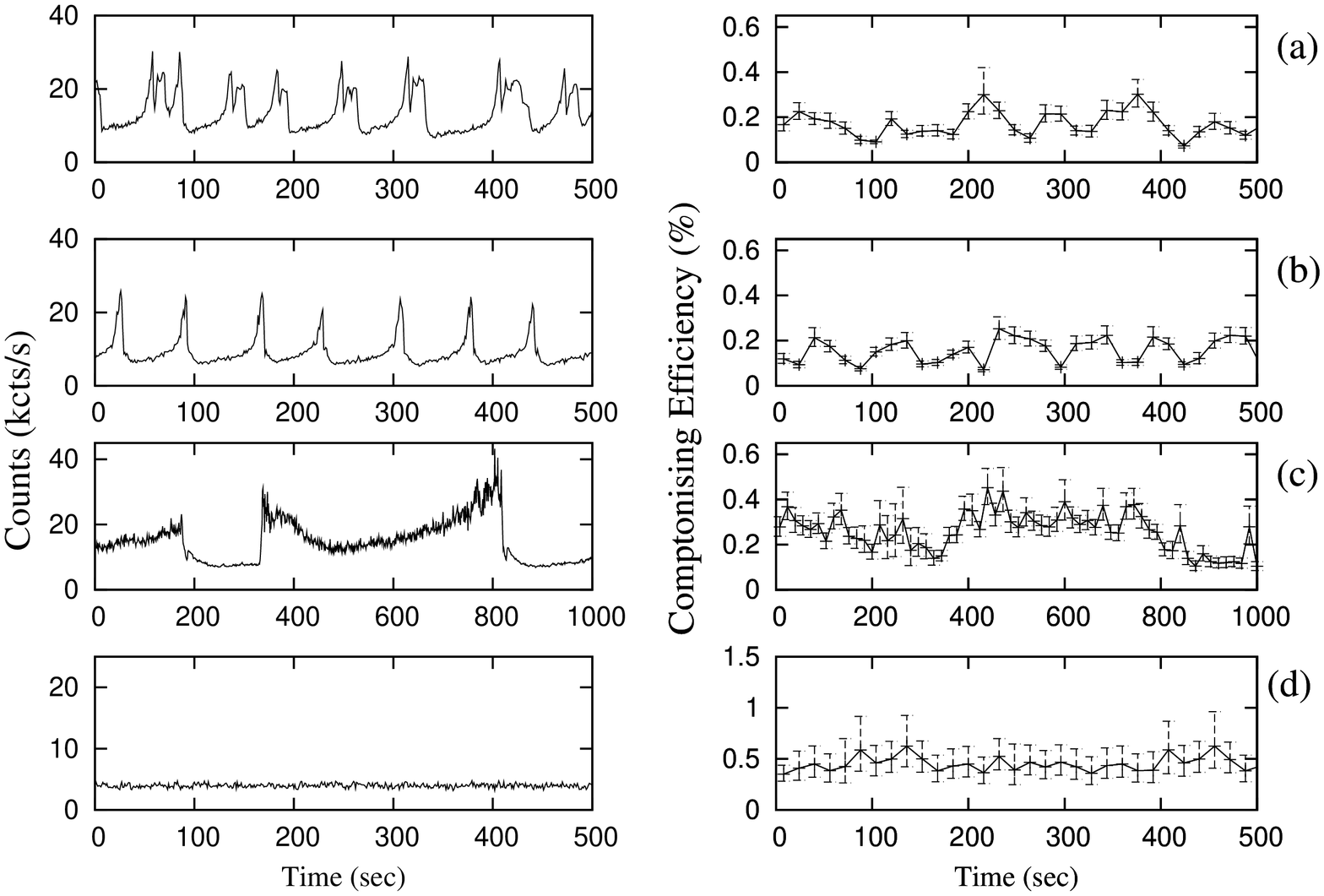,angle=270,width=10truecm}}}
\noindent{\small{\bf Fig 2:} 
Results of the analysis of (a) $\kappa$, (b) $\rho$, (c) $\theta$ and (d) $\chi$
classes are shown. Left panel: Light curve in $2.0-40.0$ keV range. Right panel:
Comptonizing Efficiency (CE) in percent, obtained from $16$s binned data.}
\end{figure}

\begin{figure}[h]
\centerline{
\vbox{
\vskip -8.0cm
\hskip 1.0cm
\psfig{file=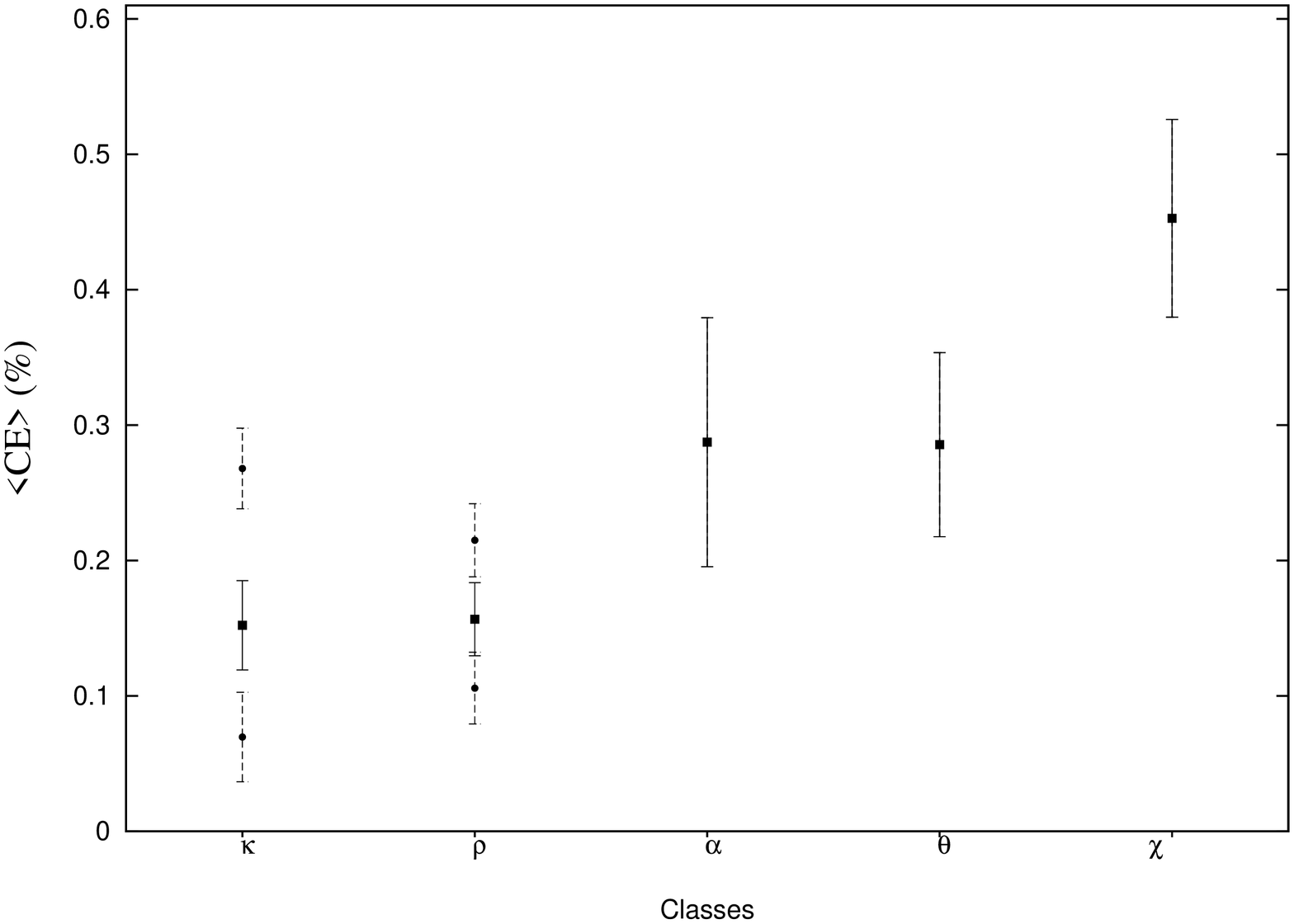,angle=270,width=8truecm}}}
\centerline{
\vbox{
\vskip -9.0cm
\hskip 1.0cm
\psfig{file=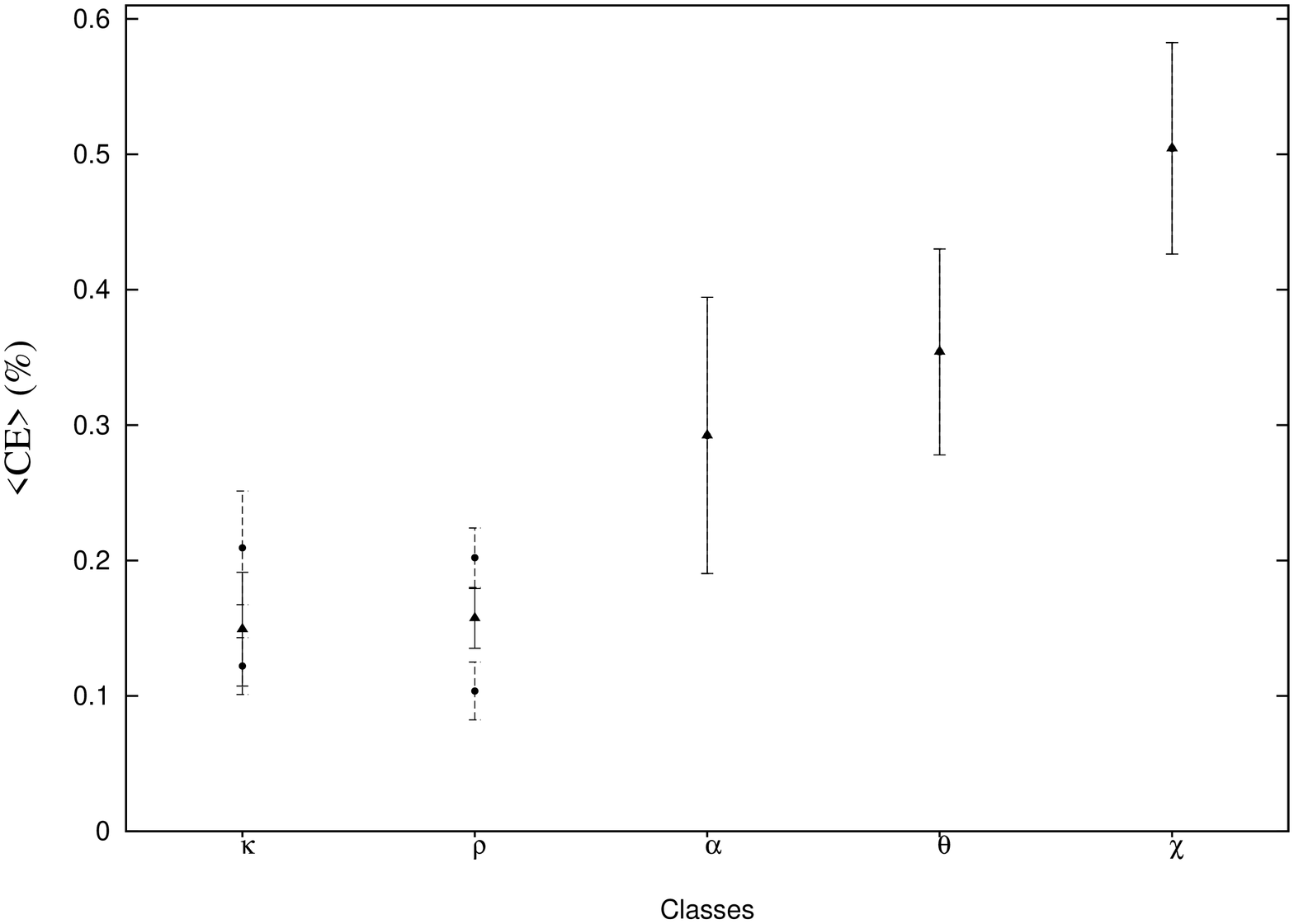,angle=270,width=8truecm}}}
\noindent{\small{\bf Fig 3} Variation of averaged Comptonizing efficiency ($<CE>$). Average values of the
two sets are shown in filled squares and filled triangles, while the average CE for a
the burst-off and burst-on states of a given class is shown in filled circles. Note that
all burst-off states have higher average CE values, while all burst-on states have lower
average CE values indicating that the Compton cloud is of larger size.}
\end{figure}

Spectral analysis for the PCA data is done by using ``standard2" mode data which
have $16$ sec time resolution.
The source spectrum is generated using FTOOLS task ``SAEXTRCT" with $16$ sec
time bin from ``standard2" data. The background fits file is generated from the ``standard2" fits file by the
FTOOLS task ``runpcabackest" with the standard FILTER file provided with the package.
The background source spectrum is generated using FTOOLS task ``SAEXTRCT" with $16$ sec time bin from
background fits file. The standard FTOOLS task ``pcarsp" is used to generate the response file with appropriate
detector information. The spectral analysis and modeling was performed using XSPEC (V.12) astrophysical fitting
package. For the model fitting of PCA spectra, we have used a systematic error of $1\%$. The
spectra are fitted with {\it diskbb} and {\it power-law}
model along with $6.0\times 10^{22} cm^{-2}$ hydrogen column absorption
(Muno et al., 1999) and we used the Gaussian for iron line as required for best fitting.
During fitting of the spectra we have adopted the technique taken by Sobczak et al. (1999), to obtain
the values of spectral parameters. We have calculated error-bars at 90\% confidence level in each case.
In soft states, the diskbb model is fitted in the $2.0 - \sim 10.0$ keV range.
and in hard states the diskbb is fitted in the $2.0 - \sim 5.0$ keV.
This upper limit changes dynamically while fitting of the spectra.

To have the spectral evolution with time for each classes, we have generated PCA spectrum
($2$ to $40$keV) with a minimum of $16$s time interval along with the background spectrum and response
matrix. This procedure is repeated with every $16$s shift in time interval since the minimum
time resolution in `standard2' data is $16$s. We use `timetrans' task to select
each and every step of time interval. 

\section{The Efficiency of Comptonization}

We first corrected for the (energy dependent) hydrogen column absorption feature
because we are interested in the number photons which were originally emitted.
We compute the photon numbers emitted in the power-law ($N_{PL}$) and the multi-color blackbody components
($N_{BB}$) and the ratio $N_{PL}$/$N_{BB}$ will be called the
Comptonization efficiency (CE).
The number of black body photons are obtained following Makishima et al., (1986)
and Comptonized photons $N_{PL}$ are calculated by using the power-law equation given below,
\begin{equation}
P(E)=N(E)^{-\alpha},
\end{equation}
where, $\alpha$ is the power-law index and $N$ is the total counts/s/keV at $1$keV.
It is reported in Titarchuk (1994), that the Comptonization spectrum will have a peak around
$3 \times T_{in}$. Thus the power law equation is integrated from $3 \times T_{in}$ to $40$keV to
calculate total number of Comptonized photons in counts/s. The presence of
line emissions are taken care of by usual Gaussian fits. Apart from the disk blackbody and power-law
model, we also tried using CompST model (Sunyaev \& Titarchuk, 1980). However, our conclusion remains the same.
In the Fig. 1, we show an example of the fitted PCA spectrum with {\it diskbb} model along with
the fitted components.  We also show the residuals to characterize the goodness of the fit.
Here we included HXETE data. However, in the rest we do not use HXETE data as
the CE is not affected by more than a percent.
The photon numbers and CE are calculated with the parameters obtained from the fitting.
The black body spectrum is simulated with $T_{in}=1.15~^{+0.057}_{-0.052}$ keV. The calculated number of
black body photons from $0.1-6.5$keV is $156.03~^{+17.10}_{-13.90}$ kcounts/s. In the same way,
the power-law spectrum is simulated with the power-law index = $2.057~^{+0.154}_{-0.147}$.

The calculated number of the Comptonized photons between $3.45-40$ keV is $0.39~^{+0.06}_{-0.05}$ kcounts/s.
The ratio between the power-law photon and the black body photon is $0.25~^{+0.08}_{-0.07}$\%.
This means that only $0.41$\% of the soft photons are Comptonized by the low angular momentum flow
component generally referred to as the Centrifugal Pressure supported Boundary Layer or CENBOL.

In Table.~2, a comparison of the fitted parameters are given with error at $90\%$ confidence level.
The first column gives the class name and the second column gives the burst-off (hard) and
the burst-on (soft) states, if present, in each class. The third column gives the black body temperature
of the fitted Keplerian disk ($T_{in}$ in keV). The fourth column gives the maximum energy in keV up to
which {\it diskbb} model is fitted and soft photon number is calculated. The fifth column
gives the count rates of the soft photons. The next column gives the index associated with the
fitted power-law component. The seventh column gives the derived hard photon counts per second.
The eighth column gives the ratio of the photon numbers of the fifth and the seventh columns. This
is the so-called Comptonizing efficiency or CE. We arranged the classes in a way that the average
CE changes monotonically. We find it to be high in harder classes and low
in the softer classes. The final column gives the value of reduced $\chi^2$ for the fits.

\begin{table*}
\begin{center}
%\begin{tabular}{crrrrrrrr}
\begin{tabular}{|c|c|c|c|c|c|c|c|c|}
\hline
\multicolumn{2}{|c|}{} & $T_{in}$ & Energy & Soft & Power law &  Hard & CE  & \\
\multicolumn{2}{|c|}{Class} &  &  &  Photon & index &  Photon &  &$\chi^2$ \\
\multicolumn{2}{|c|}{} & (keV) & keV &  (kcnt/s) &   & (kcnt/s) & (\%) & \\
\hline
$\kappa$ &h& $1.150~^{+0.057}_{-0.052}$& 6.5 & $156.03~^{+17.10}_{-13.90}$ & $2.057~^{+0.154}_{-0.147}$ &$0.39~^{+0.06}_{-0.05}$ & $0.25~^{+0.08}_{-0.07}$ & 1.2  \\ %200
\cline{2-9}
          &s& $1.900~^{+0.027}_{-0.026}$& 8.0 & $432.72~^{+13.95}_{-13.15}$ & $2.983~^{+0.243}_{-0.234}$ &$0.21~^{+0.03}_{-0.03}$ & $0.05~^{+0.010}_{-0.008}$ & 1.5  \\%248
\hline
$\rho$ &h&$1.382~^{+0.031}_{-0.029}$& 5.75 &$382.33~^{+18.42}_{-16.73}$ & $2.505~^{+0.101}_{-0.099}$ &$0.95~^{+0.09}_{-0.08}$ & $0.25~^{+0.04}_{-0.03}$ & 1.54 \\ %216
\cline{2-9}
       &s&$1.843~^{+0.022}_{-0.022}$& 8.0 &$419.92~^{+11.55}_{-10.97}$ & $2.331~^{+0.160}_{-0.155}$ &$0.34~^{+0.04}_{-0.03}$ & $0.08~^{+0.01}_{-0.01}$ & 0.99  \\ %184
\hline
$\alpha$ &-&$1.213~^{+0.082}_{-0.070}$& 4.5  &$261.36~^{+38.46}_{-28.74}$ & $2.067~^{+0.062}_{-0.061}$ &$0.81~^{+0.04}_{-0.04}$ & $0.31~^{+0.08}_{-0.07}$ & 1.01  \\ %200
\hline
$\theta$ &h&$1.391~^{+0.039}_{-0.037}$& 6.0 &$416.84~^{+25.73}_{-22.85}$ & $2.501~^{+0.089}_{-0.087}$ &$1.15~^{+0.09}_{-0.08}$ & $0.28~^{+0.05}_{-0.04}$ & 1.3  \\ %504
\cline{2-9}
         &s&$1.317~^{+0.034}_{-0.032}$& 5.5 &$388.95~^{+21.46}_{-19.19}$ & $3.248~^{+0.157}_{-0.152}$ &$0.65~^{+0.08}_{-0.07}$ & $0.17~^{+0.03}_{-0.03}$ & 1.5  \\ %200
\hline
$\chi$ &-&$1.200~^{+0.117}_{-0.094}$& 4.5 &$126.80~^{+27.74}_{-18.46}$ & $1.968~^{+0.076}_{-0.074}$ &$0.57~^{+0.03}_{-0.03}$ & $0.45~^{+0.17}_{-0.12}$ & 1.2  \\ %200
\hline
\end{tabular}
\end{center}
\caption{Parameters for the spectral fits with {\it diskbb} plus {\it powerlaw}
models for all the variability classes.}
\end{table*}

\section{Results}

So far, we have shown that our model independent CE varies from one variability class
to another. We now present the dynamical analysis of the light curves of some of the
variability classes. In Figs. 2(a-d), we show the results of $\kappa$, $\rho$,
$\theta$ and $\chi$ respectively. In each class we show two panels.
The left panel is the variation of the photon counts with time and the right
panel shows the variation of Comptonizing Efficiency (CE) calculated using $16$ seconds
of binned data. The error bar is provided at $90\%$ confidence level.
In Fig. 2a, an analysis of a $500$s chunk of the $\kappa$ class of observation
is shown. The photon counts become high $\sim 30,000$/sec and low ($\sim 10,000$/sec)
aperiodically at an interval of about $50-75$s. In the low count regions, the spectrum is harder and
the object is in the burst-off state.
The Keplerian photon varies between $150$ to $450$ kcounts/sec
and Comptonized photon varies between $0.39$ to $0.21$ kcounts/sec.
The average CE factor can vary from $\sim 0.07$ to $\sim 0.25$ depending on
whether we have burst-on or burst-off respectively.
In Fig. 2b, the blackbody photon count in the simulated spectrum varies
between $350$ to $420$ kcounts/s and Comptonized photon counts vary around $0.95$ to $0.34$ kcounts/s
Hence CE rises to a maximum of $0.25$\%.
In Fig. 2c, the analysis of a $1000$s data of the $\theta$ class is done. The $\theta$
class can be divided in two regions depending on the photon count rates.
In the soft dip region, the photon count is lower than
$10000$ counts/sec. In this region, the spectrum is softer
and the CE amount of interaction is around $0.17\%$.
In the other (hard dip) region, say between $350$s and $820$s,
the photon count is higher and vary from $10,000$ counts/sec to
$30,000$ counts/sec. In this region, CE reaches to a high value of $0.28$\%. It is believed that
the CENBOL is suddenly removed by magnetic effects and the soft dip is formed (Nandi et al. 2001).
In  Fig. 2d,  we show the results of $\chi$ class.
We note that the average Comptonizing efficiency (CE) to be $\sim 0.4-0.5$\%, the highest of
all the classes discussed so far.
We also analyzed an intermediate class called $\alpha$ (which is basically a
combination of the $\rho$ and $\alpha$ classes mentioned above. We find the CE
We analyse the data of $2000$s as this class displays a long time variation.
The CE varies between $0.05$\% to $\sim 0.6$\% with an average of $0.3$\%.

\section{Class transition due to variation in Geometry?}

In Fig. 3(a-b), we plot the average CEs of the five classes we analyzed in the previous section.
If we arrange them in the sequence of increasing average CE, we note that the sequence becomes
$\kappa \rightarrow \rho \rightarrow \alpha \rightarrow \theta \rightarrow \chi$.
We took two observations in each case and plotted both the cases, one with squares (a) and
the other with triangles (b). A further division in average CE when both the burst-on and burst-off regions
are present, is also shown with filled circles, the bottom one
being for burst-on and the top one is for burst-off.
In Papers I and II, examples were presented
of $\kappa \rightarrow \rho$, $\chi \rightarrow \rho$, $\chi \rightarrow
\theta$, $\rho \rightarrow \alpha$ transitions. In Naik et al. (2002b),  where
IXAE data was analyzed, the transitions $\rho \rightarrow \alpha \rightarrow
\chi$ were hinted at. We note that these transitions are
expected: $\kappa$ and $\rho$ are adjacent; $\chi$ and $\theta$ are
adjacent; $\rho$ can transit to $\chi$ via $\alpha$.
The class $\theta$ is an anomalous case as it is basically formed when the
Compton cloud (in our picture, CENBOL) of a $\chi$ class is abruptly removed.
Similarly, $\alpha$ is a combination of $\rho$ and $\chi$. We thus have a
natural explanation of the observed transitions in terms of the smooth variation
of the average Comptonizing efficiency,
which in turn depend of the geometry of the Compton cloud.
Since $\theta$ is an anomalous class which forms in presence of strong magnetic field only,
it may be skipped in some transition.
However, we claim that $\kappa$ cannot transit to $\chi$ directly,
for instance, without passing through $\rho$.

\section{Conclusion}

In this paper, we have analysed the classes of GRS 1915+105 which have
demonstrated class transitions. Purely in a model independent way, we
computed the ratios of the power-law photons and the soft photons
coming from the Keplerian disk in all these classes. We ignore very high energy photons
from HXETE since its contribution is less than a percent and excluding this does not change
our conclusion. Since the power-law photons are
are believed to be formed due to inverse Comptonization of the injected soft photons,
the only way the ratio can change is to change the geometry of the
Compton cloud. In CT95 model, the puffed-up part of the
low angular momentum flow (CENBOL) remains big in harder (burst-off) states
and collapses in softer  (burst-on) states. Thus we naturally see that the average
Comptonizing efficiency in burst-off states is higher than that in burst-on states.
Not only that, all the observed transitions are found to reside side-by-side in the
class vs. CE plane. In other words, for a transition, the average CE changes smoothly.
This implies that the geometry of the Compton cloud also varies from class to class.
The geometry of a Compton cloud can depend on several physical conditions. In CT95, the
CENBOL collapses when the accretion rate of the Keplerian disk is increased. This
in softer state, CE is low. When the disk rate is very low, the CENBOL cannot be cooled down.
It remains big, as in an ion-pressure supported torus (Rees et al. 1982),
and thus the interception of soft photons could be large, thereby increasing CE. We believe that
this interpretation can be used even for classes not discussed in this paper. This work
is being done and the results would be presented shortly.

Recently,  Remillard and McClintock. (2006) presented evidances that  perhaps 
GRS 1915+105 is an extremely rotating Kerr black hole. However, results discussed in this 
paper are directly the analysis of the observational data and the interpretation
of variation of the Comptonizing efficiency is totally model independent. In other words,
our conclusion of the variation of the geometry from one class to another remains 
valid even when the spin of the black hole is extreme.

The work of P. S. Pal is supported by CSIR fellowship.

{}

\end{document}